\documentclass[prl,aps,twocolumn,superscriptaddress]{revtex4}
\usepackage[dvips]{graphicx}
\usepackage{txfonts}

\begin{document}
\title{Selfgravitation  in a general-relativistic  accretion of steady fluids}
\author{Bogusz Kinasiewicz}
\affiliation{M. Smoluchowski Institute of Physics, Jagiellonian University, Reymonta 4, 30-059 Krak\'{o}w, Poland}
\author{Patryk Mach}
\affiliation{M. Smoluchowski Institute of Physics, Jagiellonian University, Reymonta 4, 30-059 Krak\'{o}w, Poland}
\author{Edward Malec}
\affiliation{M. Smoluchowski Institute of Physics, Jagiellonian University, Reymonta 4, 30-059 Krak\'{o}w, Poland}

\begin{abstract}
The selfgravity of an infalling gas can alter significantly the  accretion of gases. In the case of 
spherically symmetric steady flows  of polytropic perfect fluids
the mass accretion rate achieves maximal value when the mass  of the fluid is  1/3 of the total mass.
There are two  weakly accreting regimes, one over-abundant  and the other poor  in fluid content.
The analysis within the newtonian gravity suggests that selfgravitating fluids can be 
unstable, in contrast to the accretion of test fluids.
\end{abstract}

\maketitle

\section{introduction.}

The analysis of the spherical steady accretion of gas onto a gravitational center 
 starts with the Newtonian accretion discussed  by
Bondi \cite{bondi}. Michel \cite{michel}, Shapiro and Teukolsky \cite{shapiro_teukolsky}, 
and others \cite{das} investigated the accretion of test fluids
in the Schwarzschild space-time. Recent analysis showing the importance of backreaction is
that of Karkowski et al. \cite{Kinasiewicz}.  
In this lecture we follow the  general-relativistic description of 
selfgravitating fluids  that has been formulated in \cite{malec} and present main results 
of \cite{Kinasiewicz}.

One of the  reasons for inspecting steady flows in the full generality, when infalling gas
modifies the space-time geometry,  is the wish to see the effects of backreaction in a simple
but nontrivial accretion model. We show later, following \cite{Kinasiewicz},
that  the mass accretion rate  $\dot m$ behaves  like
$(1-x)^2\dot m_B$, where  $x=m_f/m$ is the ratio of the mass of the fluid to the total mass
and $\dot m_B$ is the mass accretion rate for the accretion of test fluids.
Since $\dot m_B$ is proportional to the fluid abundance, hence to $x$, one  
sees that there  exists a maximum of $\dot m$ when $x=1/3$.  The mass accretion rate $\dot m$ sets 
an upper limit for the luminosity of the system. This simple
analysis of accretion suggests the  existence of two  weakly luminous    regimes:
one   rich in fluid  with $m_f\approx m $  and the other with a small amount
of fluid, $m_f/m\ll 1$. It is clear from this reasoning that it is reasonable 
 to take into account selfgravitation of accreting gases whenever their mass 
is comparable to the central mass. On the other hand,  the stability properties of accreting fluids 
can depend on backreaction, however small.  
The  stability of  steady flows in 
a fixed relativistic background as well as in the newtonian context has been shown  by Moncrief 
\cite{Moncrief}. We show later that the steady newtonian acretion of selfgravitating gases 
does not exclude instability.
 
Trying to put matter in the astrophysical context, notice that
one can expect the mass of accreting fluids to be  comparable to the central mass  
only in the case of heavy and luminous compact objects;
compact galactic centers probably constitute fair examples. This suggests that
the accretion physics of such objects should include the selfgravity of accreting matter.
Let us mention here that there exists a fundamental physical aspect that the astrophysics 
of accreting matter might be able to resolve. Namely, the standard point of view is that  compact objects of
 a mass exceeding several solar masses  constitute black holes.  That interpretation
has been challenged by Mazur and Mottola \cite{Mazur} who developed 
the notion of gravatars --- stars almost 
as compact as black holes but dispossessed of event horizons. Gravastars are supposed 
to be  built from matter violating energy conditions \cite{Hawking} and 
in this way they would avoid the standard compactness
and mass limitations, like the Buchdahl theorem \cite{Buchdahl} or Chandrasekhar
type bounds on mass. There is an interesting motivation coming from the  quantum-gravity 
type arguments, that inspired the concept of gravastars, but we leave it aside. 
It is disputable  whether observations can distinguish compact bodies endowed with a hard surface
from black holes within the known nowadays accretion models. For a discussion of that issue see
 \cite{Abramowicz} and  \cite{Narayan}. We believe that the resolution of that
problem requires the inclusion of  general-relativistic aspects such as backreaction. 
In this talk we shall report the recent progress in this direction.

\section{Main equations.} 

We will consider a spherically symmetric compact ball of a fluid falling onto a non-rotating black hole.
The equations are taken from   \cite{malec}. We will use  comoving coordinates
\begin{equation}
ds^2 = -N^2 dt^2 + \alpha dr^2 + R^2 \left( d \theta^2 + \sin^2 \theta d\phi^2 \right),
\label{ds}
\end{equation}
where the lapse $N$, $\alpha$ and the areal radius $R$ are functions of a coordinate radius $r$ and
an asymptotic time variable $t$. The nonzero components of the extrinsic curvature $K_{ij}$ of
the $t = \mathrm{const}$ slices read $K_r^r = \frac{1}{2N\alpha} \partial_t \alpha, \;\;
K_\phi^\phi = K_\theta^\theta = \frac{\partial_t R}{NR} = \frac{1}{2} ( \mathrm{tr} K - K_r^r)$.
Here $\mathrm{tr} K = \frac{1}{N} \partial_t \ln \left( \sqrt{\alpha} R^2 \right)$.
The mean curvature of two-spheres of constant radius $r$, embedded in a Cauchy hypersurface is
 $k = \frac{2 \partial_r R}{R \sqrt{\alpha}}$. The energy-momentum tensor of the  perfect fluid reads
 $T^{\mu\nu} = (p + \varrho) u^\mu u^\nu +  pg^{\mu\nu}$ (where we assume the barotropic
equation of state), 
 $u_\mu u^\mu = -1$ with  $u^\mu$ denoting the four-velocity of the fluid,
 $p$ is the pressure and $\varrho$ the energy density in the comoving frame.

The areal velocity $U$ of a comoving particle labelled by coordinates $(r,t)$
 is given by $U(r,t) = \frac{\partial_t R}{N} = \frac{R}{2}(\mathrm{tr} K - K_r^r)$.
 From the Einstein constraint equations one has
\begin{equation}
kR = 2 \sqrt{1 - \frac{2m(R)}{R} + U^2}.
\label{p}
\end{equation}
Here and below we use  $\partial_r = \sqrt{\alpha}(kR/2)\partial_R$
 in order to eliminate
the differentiation with respect to the comoving radius $r$. The quasilocal mass
$m(R)$ is defined by $\partial_R m(R) = 4 \pi R^2 \varrho$. The mass
accretion rate is $\dot m = (\partial_t - (\partial_t R)\partial_R) m(R) =
- 4 \pi NR^2U (\varrho +p ) $ or (equivalently, but perhaps more
familiarly) $\dot m = -4 \pi R^2 n U$, where $n$ is the baryonic density.
We assume  a steady collapse of the  fluid, which means that all its 
characteristics are constant at a fixed areal radius $R$: 
$\partial_t X|_{R = \mathrm{const}} = (\partial_t - (\partial_t R) \partial_R) X = 0$,
 where $X = \varrho, U, a^2$. $a=\sqrt{\partial_\rho p} $ is the 
speed of sound. Strictly saying, a stationary accretion must lead to 
to the increase of the central mass and of some geometric quantities. This  
in turn means that the notion of the "steady accretion" is approximate ---
it demands that the mass accretion rate is small and  the time scale is  short,
so  that the quasilocal  mass  $m(R)$  does not change significantly.
Thence also geometric quantities, like the mean curvature $k$ or 
the area of the black hole, remain practically constant.

 The Einstein evolution equation
 $\partial_t U = k^2 R^2 \partial_RN/4 - m(R)N/R^2 - 4 \pi N R p$ and
 the energy conservation equation $\partial_t \varrho = -N \mathrm{tr} K (\varrho + p)$
constitute (assuming steady flow) ordinary differential equations with respect to $R$.
  
   It is easy to show that now the mass accretion rate is constant,
 $\partial_R \dot m = 0$ \cite{malec}.

By a black hole is understood a region within an apparent horizon to the future, i.e.,
a region enclosed by an outermost sphere $S_A$ on which the optical scalar $\theta_+
\equiv \frac{pR}{2} + U$ vanishes \cite{nom}. (The other condition, that
$\theta_-\equiv \frac{pR}{2} - U > 0$ for all spheres outside $S_A$, is
satisfied trivially for a steady accretion.) That means that on $S_A$
the ratio $2m_{BH}/R_{AH}$ becomes 1, where $R_{AH}$ is the areal radius
of the apparent horizon and $m_{BH}\equiv m(R_{AH})$ is the mass of the black hole.
    We assume the polytropic equation of state
   $p = K \varrho^\Gamma$, with $\Gamma$ being a constant, ($1 < \Gamma \le 5/3$).

Furthermore, we assume that the radius $R_\infty$ of the ball of fluid and other
boundary data are such  $|U_{\infty}| \ll \frac{m(R_\infty)}{R_\infty}
 \ll a_\infty$. These boundary conditions allow one to join smoothly
the internal solution  (the steady fluid) with the external Schwarzschild geometry.
 The momentum conservation equation $\nabla_\mu T^\mu_r = 0$ ---
in comoving coordinates
 \begin{equation}
\label{euler}
N \partial_R p + \left( p + \varrho \right) \partial_R N = 0
\end{equation}
--- can be integrated, yielding (with $N(R_\infty) = 1$)
\begin{equation}
a^2 = - \Gamma + \frac{\Gamma +a^2_\infty}{N^\kappa}.
\label{bernoulli}
\end{equation}
Here $\kappa = \frac{\Gamma - 1}{\Gamma}$. Equation  (\ref{bernoulli}) is 
 the general-relativistic version of the Bernoulli equation. 
Equation (\ref{euler}) implies also that the lapse expresses as follows
$N=\frac{Cn}{p + \varrho}$, with  the baryonic number density
 $ n = C \exp \int_{\varrho_\infty}^\varrho ds \frac{1}{s + p\left( s\right) }. $
The baryonic current is conserved, $\nabla_\mu \left( n u^\mu \right) = 0$.
  
The sonic point is  as a location where
the length of the spatial velocity vector $|{\vec U}| = |U|/(kR/2)$ equals $a$.
 Therefore   $|U| = \frac{1}{2}kRa$. In the Newtonian limit this
  coincides with the standard requirement $|U|=a$. In the following we will denote
   by the asterisk all values referring to the sonic points, e.g. $a_\ast$, $U_\ast $, etc.
   The four characteristics, $a_\ast$, $U_\ast$, $m_\ast / R_\ast$ and $c_\ast$ are related
   \cite{malec}, $a_\ast^2 \left( 1 - \frac{3m_\ast}{2R_\ast} + c_\ast \right) = U_\ast^2 =
   \frac{m_\ast}{2R_\ast} + c_\ast$, where $c_\ast = 2 \pi R^2_\ast p_\ast$. The infall velocity $U$ reads
\begin{equation}
U=U_\ast \frac{R^2_\ast}{R^2}
\left(  \frac{1 + \frac{\Gamma}{a^2}}{1 + \frac{\Gamma}{a^2_\ast} } \right)^{1/(\Gamma -1)}.
\label{U}
\end{equation}
Here $U_\ast$ is the negative square root. From the relation between the pressure and the
energy density, one obtains, using  equation (\ref{bernoulli})
\begin{equation}
\varrho = \varrho_{\infty } \left( a/a_\infty \right)^{2/(\Gamma - 1)} = \varrho_\infty
\left( - \frac{\Gamma}{a_\infty^2} + \frac{\frac{\Gamma}{a_\infty^2}  + 1}{N^\kappa} \right) ^ \frac{1}{\Gamma - 1},
\label{rho}
\end{equation}
where the constant  $\varrho_\infty$ is equal to the mass density of a collapsing
fluid at the boundary $R_\infty$. The steady fluid is described by equations
(\ref{bernoulli} -- \ref{mR}). They constitute an integro-algebraic system of equations,
with a bifurcation point at the sonic point, where two branches (identified as an accretion
or a wind) do cross; that is a well known feature of that problem, present also in models
 with a test fluid \cite{bondi} -- \cite{das}, \cite{malec}. That requires some caution in doing
 numerics and a careful selection of the solution branch. Notice that these equations
  are expressed exclusively in terms of quantities that are steady in the sense of the former definition.
 
\section{Calculation of the mass accretion rate.}

    In numerical calculations it will be convenient to represent the mass $m(R)$ in the form
\begin{equation}
m(R) = m - 4 \pi \int_R^{R_\infty} dr r^2 \varrho.
\label{mR}
\end{equation}
Notice that $m(\infty) = m_\infty \approx m$. For a steady  flow
one obtains \cite{malec}
\begin{eqnarray}
N & = & \frac{kR}{k(R_\infty) R_\infty} \beta(R),
\nonumber\\
\beta (R) & = &  \exp \left( - 16 \pi \int_R^{R_\infty} \left(  
 p+ \varrho \right) \frac{ds}{k^2 s} \right).
\label{Nb}
\end{eqnarray}
The full description of the steady accretion  is
 given by equations (\ref{p}), (\ref{mR}), (\ref{bernoulli}), (\ref{U}) and 
 (\ref{Nb}), which are equivalent  to the former set of equations.

In the forthcoming part of the talk we shall describe main points of the 
derivation of the accretion mass rate, following the presentation of Karkowski
et al. \cite{Kinasiewicz}. 

In the first step, one shows in \cite{Kinasiewicz},
 that outside the sonic horizon
solutions of the equations with backreaction can be estimated
by relevant solutions describing the   stationary accretion of test fluids.
We supplement the foregoing list of conditions by
another, that the mass $m_f$
of the fluid outside the sonic horizon is of the order of the mass
$m_\ast $  within the sonic horizon;   
That allows one to show that the function $\beta $ appearing 
in equation (\ref{Nb}) is essentially 1 and
$c_\ast = 2 \pi R^2_\ast p_\ast \ll a^2_\ast $ can be ignored.  As 
a consequence one infers that in the case of stationary accretion
of selfgravitating fluids some  characteristics of the sonic point
--- $a^2_{\ast } $, $U^2_\ast $ and $m_\ast /R_\ast $ --- are the same as
for the test fluid accretion, provided that asymptotic data $\rho_\infty $ and
$a^2_\infty $ are identical \cite{Kinasiewicz}. In particular, in both cases 
characteristics of the sonic point do not depend on the mass --- either the total
mass $m$ or the mass $m_\ast $ of the accretor. 

In the second step one shows that it is possible to estimate the energy 
density $\varrho $ of a system with backreaction using information about
solutions in the test fluid approximation. It is done in  \cite{Kinasiewicz}. 
We shall describe  in a  more detail how one obtains the desired form of
 estimates onto the energy density of test fluids.  
One starts with the  estimate on the inverse of the lapse 
function $N\left(\delta \right) =\sqrt{1-\delta }$, where
$\delta \equiv  {2m\over R} -U^2 $ is less than one (and in fact strictly smaller than
$3/4$).  It is easy to show, applying the method
of differential inequalities,  that $N^{-1}\left( \delta \right) \le 
1+ \delta /\left( 2N^3\left( \delta \right) \right) $. Since $\delta \le 2m/R$,
one has $N^{-1}\left( \delta \right) \le  1+ m /\left( RN^3\left( \delta \right) \right) $.
Notice now that the lapse is an increasing function and $N(R) \ge N_\ast
\equiv N(R_\ast )$ for $R\ge R_\ast $. Therefore one arrives at the inequality 
\begin{eqnarray}
&&N^{-1}\le 1 +{1\over N_\ast^{3}}\times {m\over R}.
\label{lapse1}
\end{eqnarray}
Having that, one can derive a bound onto $\varrho $. 
One obtains, applying the inequality (\ref{lapse1}),
\begin{equation}
\varrho (R) \le \varrho_\infty \left( 1+ \left( \Gamma -1\right)
 {1+{\Gamma \over a_\infty^2} \over  \Gamma N^3_\ast  }
\times {m \over    R}\right)^{1/\left( \Gamma -1 \right) } .
\label{r1}
\end{equation}
 The lapse at the sonic point $N_\ast =1/\sqrt{1+3a^2_\ast }$ and $m/R_\ast = a^2_\ast N^2_\ast$;
thus a constant $\alpha \equiv 
 {1+{\Gamma \over a_\infty^2} \over  \Gamma N^3_\ast  }
 {m \over    R_\ast} $ can be written as
\begin{eqnarray}
\alpha  
&=& {1\over \Gamma }\sqrt{1+3a^2_\ast }
\left( a^2_\ast +{\Gamma a^2_\ast \over a^2_\infty }\right)  .
\label{alpha}
\end{eqnarray}
One can prove following bounds (see equation (5.2) in \cite{malec})
\begin{equation}
{a^2_\ast \over a^2_\infty}\le  \frac{2}{5 - 3 \Gamma + \frac{3  a^2_\ast 
(\Gamma - 1)^2 (9\Gamma - 7)}{4 \Gamma \left(1 + 3  a^2_\ast  \right)}}.
\label{theorem2}
\end{equation}
This  implies a series of  estimates:
i) for $1<\Gamma \le 4/3$ one has ${a^2_\ast \over a^2_\infty} <2$;
ii) in the range $4/3<\Gamma \le 3/2$ holds ${a^2_\ast \over a^2_\infty} <4$;  
iii) when $\Gamma >3/2$ and $a^2_\ast <0.2$ one obtains 
 ${a^2_\ast \over a^2_\infty} < 2.3/a^2_\ast $ and $a_\ast^2 \le 1.6a_\infty $,
iv) while  for $\Gamma >3/2$ and $a^2_\ast \ge 0.2$
it follows ${a^2_\ast \over a^2_\infty} <4$.
Now, if we demand that the sonic point is outside
 of an apparent horizon then $m/\left( 2R_\ast \right) <1/4 $
and that implies $a^2_\ast \le 1$. Using this fact, one can derive bounds onto the constant 
$\alpha $. Namely, in each of the above cases we have  
i)  $1<\Gamma \le 4/3\rightarrow \alpha <6$;
ii)  $4/3<\Gamma \le 3/2\rightarrow \alpha  <9.5$;  
iii)  $\Gamma >3/2$ and $a^2_\ast <0.2$:  
 $\alpha <1 + 2/a_\infty  $ 
iv)  $\Gamma >3/2$ and $a^2_\ast \ge 0.2$:
  $\alpha <8.3$.

With these estimates one obtains simple bounds onto $\rho $. We shall consider in 
more detail two of them corresponding to the cases i) and iii). 
For $\Gamma \le 4/3$ equation (\ref{r1}) can be written, 
using the inequality $\left( 1+A/x\right)^x\le e^A  $ and $\alpha \le 6$, as
\begin{eqnarray}
&&\varrho (R)\le \varrho_\infty \exp \left(   { 6 R_\ast \over R}\right) .
\nonumber\\
\label{r2}
\end{eqnarray}
In the case iii) one notices that $(1+A)^{1\over \Gamma -1}
\le 1+2A+A^2$. That leads to the estimate
\begin{eqnarray}
{ \varrho \over \varrho_\infty }&\le &
1+ 2\left( \Gamma -1\right) \alpha {R_\ast \over R}+
\nonumber\\
&&\left( \Gamma -1\right)^2 \alpha^2 { R_\ast^2\over R^2}.
\label{r3}
\end{eqnarray}
one can check the selfconsistency of the calculation
--- for instance that the function $\beta (R)\approx 1$ for $R\ge R_\ast $.
 Indeed, the exponent in the exponential funtion
$\beta \left( R_\ast \right) $ is of the order of $M_f\equiv 
4\pi \int_{R_\ast}^{R_\infty } dr r\rho$. This latter expressions reads in the case i)
\begin{eqnarray}
M_f&=&2\pi \rho_\infty R_\infty^2 \Biggl[ e^{6R_\ast /R_\infty } - 
e^6{R_\ast^2 \over R_\infty^2 }+\nonumber\\
&&
{6R_\ast \over R_\infty }e^{6R_\ast /R_\infty } - 6e^6{R_\ast^3 \over R_\infty^3 }
+\nonumber\\
&&\left( {36R_\ast^2 \over R_\infty^2 }e^{6R_\ast \over R_\infty } -
 36e^6{R_\ast^4 \over R_\infty^4 }\right) \int_1^{R_\infty /
\left( 6R_\ast \right) }dx{e^{1/x}\over x}\Biggr] .
\nonumber\\
\end{eqnarray}

Since $R_\infty \gg R_\ast $ one concludes that $M_f \approx 2\pi \rho_\infty R^2_\infty $
 which in turn is clearly much smaller than $m/R_\ast $. Thus the exponent of $\beta (R)$
is negligibly small for $R\ge R_\ast $ and  $\beta (R) \approx 1$ as desired.
A similar reasoning in the case iii) leads to $M_f \approx 2\pi \rho_\infty R^2_\infty  \left(1+
O\left( \alpha R_\ast /R_\infty \right) \right) $.
 Now $\alpha R_\ast /R_\infty \le R_\ast /R_\infty
+2R_\ast /\left( R_\infty a_\infty \right) $. 
It is easy to see, taking into account  the  boundary condition
 $a^2_\infty \gg m/R_\infty $ and $a_\ast \approx a_\infty $,
that the second  term  in $M_f$ is much smaller than 1 and therefore the expression
$O\left(  \alpha R_\ast /R_\infty \right) $
is negligible small. Thus again   $M_f\approx m_f /R_\infty \ll m/R_\infty \ll 1$,
 where $m_f$ is the mass of fluid outside
the sonic horizon; therefore one again arrives at
  $\beta \left( R_\ast \right) \approx 1$.

In the third step one shows --- using the same reasoning as in the preceding
paragraghs --- that the mass $m_f = \gamma \varrho_\infty $, where $\gamma $ is
a constant. This
is because  the mass density changes moderately outside the  sonic point, as shown by 
estimates (\ref{r2}-\ref{r3}). The calculation is lengthy but simple and we omit it here.
Taking into account $m=m_\ast +m_f$, one infers  $m_\ast / m =
 1 - \varrho_\infty (\gamma / m)$;  
the ratio $x\equiv m_\ast / m$   linearly decreases with  the growth of $\varrho_\infty$.

The  rate of  the  mass accretion $\dot m$   can be expressed as below (see equation (6.1) in \cite{malec})
\begin{equation}
\dot m = \pi m_\ast^2 \varrho_\infty \frac{R^2_\ast}{m_\ast^2} \left( \frac{a_\ast^2}{a_\infty^2} 
\right)^\frac{(5 - 3 \Gamma)}{2(\Gamma - 1)} 
\left(1 + \frac{a^2_\ast}{\Gamma } \right) \frac{1 + 3 a_\ast^2}{a^3_\infty}.
\label{dotm}
\end{equation}
The whole dependence on $\varrho_\infty$ is contained in  the factor $m_\ast^2 \varrho_\infty
 =m^3x\left( 1-x\right)^2/\gamma $.
Thus one can write 
\begin{equation}
\dot m =  m^3x{ \left( 1 -x\right)^2\over \gamma }
\pi  \frac{R^2_\ast}{m_\ast^2} \left( \frac{a_\ast^2}{a_\infty^2} 
\right)^\frac{(5 - 3 \Gamma)}{2(\Gamma - 1)} 
\left(1 + \frac{a^2_\ast}{\Gamma } \right) \frac{1 + 3 a_\ast^2}{a^3_\infty}.
\label{dotm1}
\end{equation}
This expression clearly demonstrates that the mass accretion rate  
 achieves  a maximum at $m_\ast = 2 m / 3$ and tends to zero when 
i) $m_\ast \to m$ (when the density $\varrho$ tends to zero) and ii) $m_\ast / m \to 0$. 
 
  Thus, one concludes that  \textbf{Amongst steadily accreting systems 
  of the same  $R_\infty$, $a_\infty$ and $m$ those will be most efficient for which $m_\ast = 2 m / 3$.}
   The factor 2/3 is universal --- independent of the parameters $R_\infty$, $\Gamma$ and $a_\infty$.

Since the mass of the fluid  $m_f$ is close  to $m-m_\ast $  and $m=m_f+m_{BH}$, the above
means that the maximum of the mass accretion takes place when the mass of the fluid is
a half of the mass of the black hole.

Finally let us mention that in the case  of polytropes with   $ = K n^\Gamma$
 the Bernoulli equation  reads $N \left( \Gamma - 1 - a^2_\infty \right) = \Gamma - 1 -a^2$.
The analytic arguments of the type presented above can be applied and conclusions are the same.
   Similar results should be true for  
  massive fluids in the newtonian limit  provided that the polytropic index   $\Gamma $
is clearly smaller than $5/3$.

\section{Numerical results.}

In this Section we present a small sample of the numerical investigation,
 different from that shown in \cite{Kinasiewicz}. 
The main lesson that one can infer from these is that  boundary conditions displayed in Sec.
 2 can be significantly weakened. It is clear from considerations around 
$\left( \ref{theorem2}\right) $ that the point where one can expect difficulties is
when the adiabatic index $\Gamma =5/3$, since in this case the relevant bounds are relatively 
poor. But the numerical data, that are displayed in the forthcoming Table and figures,
demonstrate    suprisingly good agreement with the statements of the preceding section.

\begin{table}[ht]
\caption{Characteristics of the sonic point $a_\ast^2$, $|U_\ast|$, $R_\ast$ for the $\Gamma
 \approx 5/3$ polytrope and $a_\infty^2 = 0.1$. The last column shows the areal radius of 
 the apparent horizon. The mass accretion rate (second column) achieves maximum at $m_{BH} \approx 0.663$.}
\begin{tabular}{cccccc}
$m_f / m$ & $\dot m$ & $R_\ast$ & $U_\ast$ & $a_\ast^2$ & $R_{AH}$ \\
\hline
\hline
$4.18\times 10^{-32}$   & $-2.75\times 10^{-48}$  & $2.76$ & $0.424584$ & $0.397394$ & $1.99$ \\ \hline
$4.18\times 10^{-22}$   & $-2.75\times 10^{-38}$  & $2.76$ & $0.424584$ & $0.397394$ & $1.99$ \\ \hline
$4.18\times 10^{-12}$   & $-2.75\times 10^{-28}$  & $2.76$ & $0.424584$ & $0.397394$ & $1.99$ \\ \hline
$0.038$                 & $-2.53\times 10^{-18}$  & $2.64$ & $0.424584$ & $0.397395$ & $1.91$ \\ \hline
$0.075$                 & $-4.62\times 10^{-18}$  & $2.53$ & $0.424621$ & $0.397703$ & $1.82$ \\ \hline
$0.12$                  & $-6.31\times 10^{-18}$  & $2.41$ & $0.424621$ & $0.397702$ & $1.74$ \\ \hline
$0.16$                  & $-7.63\times 10^{-18}$  & $2.30$ & $0.424575$ & $0.397359$ & $1.66$ \\ \hline
$0.2$                   & $-8.60\times 10^{-18}$  & $2.18$ & $0.424621$ & $0.397702$ & $1.57$ \\ \hline
$0.25$                  & $-9.26\times 10^{-18}$  & $2.06$ & $0.424621$ & $0.397703$ & $1.49$ \\ \hline
$0.29$                  & $-9.63\times 10^{-18}$  & $1.95$ & $0.424621$ & $0.397719$ & $1.40$ \\ \hline
$0.33$                  & $-9.73\times 10^{-18}$  & $1.83$ & $0.424619$ & $0.397784$ & $1.32$ \\ \hline
$0.38$                  & $-9.62\times 10^{-18}$  & $1.72$ & $0.424585$ & $0.397395$ & $1.24$ \\ \hline
$0.41$                  & $-9.30\times 10^{-18}$  & $1.60$ & $0.424611$ & $0.397586$ & $1.16$ \\ \hline
$0.83$                  & $-1.45\times 10^{-18}$  & $0.44$ & $0.424621$ & $0.397707$ & $0.32$ \\ \hline
$0.88$                  & $-8.37\times 10^{-19}$  & $0.33$ & $0.424621$ & $0.397707$ & $0.24$ \\ \hline
$0.94$                  & $-2.05\times 10^{-19}$ & $0.16$ & $0.424621$ & $0.397704$ & $0.11$ \\ \hline
\end{tabular}
\end{table}
It appears that the characteristics $a_\ast^2$, $U_\ast^2$, $m_\ast/R_\ast$ of the sonic point 
{\it practically} do not depend, for a given $\Gamma$ and $a^2_\infty$, on the energy density 
$\varrho_\infty$ (exemplary results are shown in Table 1), suggesting a precision of $10^{-5}$. 
The more reliable measure of the numerical error is the difference between the mass $m = 1$ 
(assumed in the equations) and the mass found numerically. Our results (see Table ) suggest
that numerical error is smaller than 0.5 \%. The quantity $c_\ast$ is negligible in comparison 
to other sonic point parameters. The mass $m_\ast$ behaves like $m_\ast = 1 - 
\gamma \varrho_\infty$ (Fig. 1). And finally, the extremum of $\dot m$ is achieved when 
$m_\ast \approx 2/3$ (Fig. 2). Notably, numerical results extend the regime in which the 
universality is observed into fluids characterized by the speed of sound $a_\infty$ exceeding 1. 
This is remarkable, since that implies $R_\ast < R_{AH}$ and the theoretical analysis 
presented above would not be valid.
\begin{figure}[h]
\includegraphics[width=8cm]{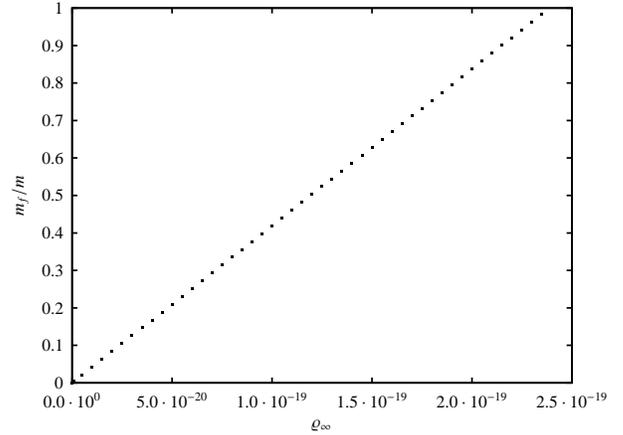}
\caption{The dependence of $\frac{m_f}{m} \approx 1 - m_\ast$ on the asymptotic mass density.}
\end{figure}
\begin{figure}[h]
\includegraphics[width=8cm]{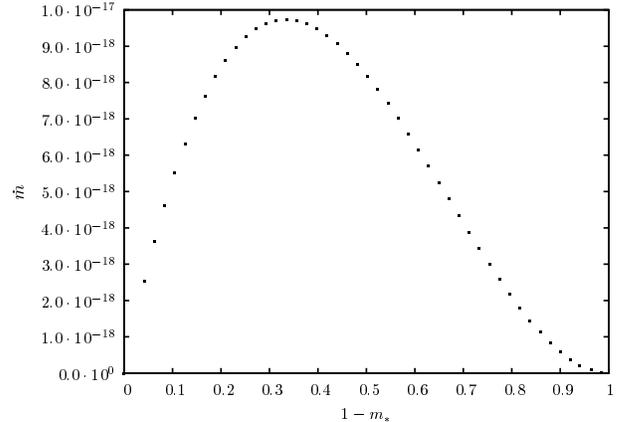}
\caption{The dependence of $\dot m$ on the ratio $\left(m - m_\ast \right) / m$.}
\end{figure}

Formula (\ref{dotm}) implies that the mass accretion rate  in the case with backreaction
(when $m_\ast < m$) can be only smaller from the mass accretion rate without backreaction 
(when $m_\ast = m$), with the same asymptotic parameters $\rho_\infty$, $a_\infty$, $m$.
Numerical results agree with the analytic conclusion
 that the proportionality coefficient  $\gamma $ depends
essentially only  on the size of support.

\section{Stability of selfgravitating fluids in lagrangian coordinates.}

The effect of selfgravitation of the steadily accreting gases on the stability
has not been studied, up to our knowledge, even in the Newtonian case.
In this section we shall fill this gap by analyzing nonrelativistic accretion.
The relativistic accretion will be done elsewhere.
The forthcoming analysis holds true for polytropes and, more generally, for 
barotropes. The newtonian equations for selfgravitating fluids in terms of
the  Eulerian time $t$ and the areal radius
$R$ read
 \begin{eqnarray}
 \partial_t U +U\partial_RU&=& -{\partial_Rp\over \varrho } - {m(R)\over R^2}
\nonumber\\
\partial_t \varrho &=&
 -{1\over R^2} \partial_R \left( R^2\varrho U\right) .
\label{Euler}
\end{eqnarray}
The stability analysis of the Bondi accretion in the Lagrange formulation
has been performed by Balazs \cite{Balazs}. His approach is correct 
but inconclusive, because of a too stringent understanding of the notion 
of linearized stability.  We show that using lagrangian variables
 one exactly reproduces the
stability result already known from the Eulerian approach \cite{Moncrief}.

  As the primary quantity 
one takes the deviation $\zeta (t,r)\equiv \Delta R(r,t)$  from  the particle position in the 
unperturbed flow. The perturbed velocity reads $\Delta U=\partial_t^L\zeta \equiv 
(\partial_t +U\partial_R )\zeta $; here  $\partial_t^L\equiv \partial_t+U\partial_R$ 
is the derivative with respect the Lagrangian time. 
The lagrangian perturbation   $\Delta \varrho $ can be related to $\zeta $
by demanding that $\Delta m(R(r=\mathrm{const} ))=0$; that implies  
\begin{equation}
\Delta \varrho =-\rho \left( {2\zeta \over R} +\partial_R\zeta \right) .
\label{rhol}
\end{equation}
 
  The   $U, a, \varrho $
solves  Eqs. (\ref{Euler}) with the mass function  $m(R)$ including the contribution of the fluid itself.
One derives from the first of Eqs. (\ref{Euler}) the equation ruling $\zeta $
(take notice during  the derivation that $\Delta m(R(r))=0$)
\begin{eqnarray}
\left( \partial_t^L\right)^2 \zeta &=&{2m(R)\zeta \over R^3} +
{1\over \varrho }\partial_R \left( a^2 \varrho \left( \partial_R\zeta 
+{2\zeta \over R }\right) \right) -
\nonumber\\
&& {2\zeta \over \varrho }\partial_Rp   
\label{zeta1}
\end{eqnarray}
The  perturbation of the other equation in (\ref{Euler})
is satisfied identically.  

Balazs tried to find solutions in the standard way,
 of the form $\zeta \left( R(r)),t\right) = \exp^{i\omega t} 
\zeta (R(r)) $ with positive $\omega^2$, that is with nonincreasing in time 
modulus; this clearly cannot be done, as shown below.  
The proper approach (adopted in the Eulerian variables by Moncrief \cite{Moncrief})
is to define an  energy of perturbed modes  by the  expression
\begin{eqnarray}
E&=&\int_VdV \varrho \Biggl( {1\over 2}\left( \partial_t\zeta \right)^2
+ {1\over 2}\left(\partial_R\zeta \right)^2 \left( a^2-U^2\right) +\nonumber\\
&&  {\zeta^2\over R^2}\left( a^2-{m\over R}-R\partial_Ra^2\right) \Biggr) ,  
\label{energy11}
\end{eqnarray}
where $V$  is  an annulus  between radii $R_\ast $ and $R_\infty $.
The interesting and nontrivial question (since the last integrand in (
\ref{energy11}) can be negative in the vicinity of a sonic point) is whether 
the energy $E$ of  (\ref{energy11}) is positive. It is enough to consider the 
region outside of the sonic horizon, because perturbations that exist
 inside will be washed away to  the black hole.
 
In order to  answer that question, notice  an  identity  
\begin{eqnarray}
&&\varrho \left( a^2-{m\over R}-R\partial_Ra^2\right) =
-\partial_R\left[  \varrho R \left( a^2 -{m\over 2R}\right) \right] 
+
\nonumber\\
&&  {2\varrho \over a^2-U^2}\left( a^2 -{m\over 2R}\right)^2  -2\pi R^2
  \varrho ;
\label{ident1}
\end{eqnarray}
notice the occurence of the negative  term at the end of the formula.
Inserting (\ref{ident1}) into the last term in the integrand of (\ref{energy11})
and integrating by parts, one obtains after lengthy calculation  that
\begin{eqnarray}
E&=&\tilde E - \left[ 4\pi R \zeta^2\rho  \left( a^2 -{m\over 2R}\right) 
\right]^{R_\infty }_{R_\ast } 
\label{energy12}
\end{eqnarray}
where 
\begin{eqnarray}
\tilde E&=&{1\over 2}\int_VdV\left( X^2+Y^2 \right) -
\nonumber\\
&&
2\pi \int_V dV \zeta^2 \varrho^2 ,
\label{energy13}
\end{eqnarray}
is not definite,
while  $X$ and $Y$ are given by 
\begin{eqnarray}
X & = & \sqrt{\varrho} \partial_t \zeta, \\
Y & = & \sqrt{\varrho} \left( \frac{2 a^2 R - m}{R^2 \sqrt{a^2 - U^2}} \zeta + \sqrt{a^2 - U^2} \partial_R \zeta \right).
\end{eqnarray}
One can easily show  that boundary  terms of (\ref{energy12}) are
 nonnegative. The  occurence of a negative term in (\ref{energy13}) means that  
 the energy $E$ of modes perturbing the selfgravitating flow may be negative.

The calculation of the derivative of the energy  with respect to the Eulerian time
yields
\begin{eqnarray}
&&\partial_tE=\int_V dV \zeta^2 {-\partial_tm(R)\over R^3}
  +
\nonumber\\
&&
4\pi  \left[ r^2\rho \left( \partial_t\zeta
\partial_R\zeta \left( a^2-U^2\right) -U\left(\partial_t\zeta \right)^2\right) 
\right]^{R_\infty }_{R_\ast } .
\label{flux1}
\end{eqnarray}
 A careful analysis of  boundary terms shows that the contribution from $R_\infty $
and $R_\ast $  is  negative definite. We can conclude that the   outflow 
through the inner and outer boundaries is nonpositive and the energy $E$ of perturbations 
cannot grow for the critical flow. Despite this fact, there is a possibility
for unstable behaviour of the solutions of the time-dependent hydrodynamic equations,
because the energy $E$ is not positive definite.

 The situation  for subsonic flows is unclear even for test fluids,
 as pointed by Garlick
\cite{Garlick}. While the analysis at $R_\infty $ yields the  same result, that
the outflow diminishes the energy $E$, one has to assume 
that the  inner outflow is completely absorbed by the surface of an attracting body, that is
$\zeta (R_1,t)=0$, in order to  conclude that   $\partial_tE \le 0$.
 For more realistic cases, with the occurence of a reflection from the inner boundary, 
the energy of the perturbation may grow, signalling an instability, in addition to the
"bulk-type instability" suggested by the above analysis.  

 It is necessary to say in this place that the 
 absence of  exponentially growing linear  modes does not mean that perturbed
solutions of the nonlinear equation will be close to the background solution
at all times. Linearized stability of a solution would imply only that evolving
 perturbations can be bounded by initial perturbations in a suitable sense.
 The existence of exponential linear modes in turn does not necessarily yields  that 
perturbations grow indefinitely. Linear instability  means that some evolving  
nonlinear perturbations   are not bounded by initial perturbations; the "strength"
of evolving perturbations does not depend on the "strength" of initial pertrbations
\cite{Ioss}.

 Restricting our attention to test fluids, notice the following.
The energy  $E$ is equal to $\tilde E$, and $\tilde E$ differs from the expression (\ref{energy13})
 only by the absence of the last,
nonpositive integral, hence it is clearly positive. The time propagation in turn is 
like in (\ref{flux1}), but without the  integral  $\zeta^2 {-\partial_tm(R)\over R^3}
=\zeta^2 {4\pi U\rho \over R}$ on the right hand side. Thus $E$ cannot grow, which excludes 
the exponential growth of $X$ and $Y$ and the long-term exponential growth
of linear modes $\zeta $. That does not guarantee that $|\zeta \left( R\right) |$ is 
time-independent. 
  
This paper has been partially supported by the MNII grant 1PO3B 01229.

\end{document}